\documentclass[letterpaper]{article}

\usepackage[margin=0.5in]{geometry}
\usepackage{microtype}
\usepackage{graphicx}
\usepackage{array}
\usepackage{booktabs}
\usepackage{makecell}
\usepackage{multirow}
\usepackage{adjustbox}
\usepackage{rotating}
\usepackage{placeins}
\usepackage{xcolor}
\usepackage{soul}
\usepackage[numbers,sort&compress]{natbib}

\usepackage{amsmath, amssymb, amsfonts, bm}

\usepackage{subfigure}

\usepackage{times}

\usepackage[numbers,sort&compress]{natbib}

\usepackage[breaklinks=true,hidelinks]{hyperref}

\begin{document}

\vspace{1 cm}

\title{Residual Gaze Behavior During Navigation in Blindness and Low Vision}

%

\author{Junchi Feng$^{a,b}$,
        Fernanda Garcia-Piña$^{c}$, 
        Mahya Beheshti$^{d}$,
        Todd E Hudson$^d$, 
        William Seiple$^c$,
        John-Ross~Rizzo$^{a,d}$
}


\maketitle

\textbf{Affiliations:}
\begin{itemize}
    \renewcommand{\labelitemi}{}
    \item $^a$ Department of Biomedical Engineering, Tandon School of Engineering, New York University, Brooklyn, NY, 11201, USA
    \item $^b$ Center for Urban Science and Progress, Tandon School of Engineering, New York University, Brooklyn, NY, 11201, USA
    \item $^c$ Lighthouse Guild, New York, NY 10023, USA
    \item $^d$ Department of Rehabilitation Medicine, NYU Grossman School of Medicine, New York, NY, 10016, USA
\end{itemize}

\clearpage

\textbf{\Huge Residual Gaze Behavior During Navigation in Blindness and Low Vision}

%


\maketitle

\begin{abstract}
\textbf{Background} Outdoor navigation poses significant challenges for people with blindness or low vision, yet the role of gaze behavior in supporting mobility remains underexplored. Fully sighted individuals typically adopt consistent scanning strategies, whereas those with visual impairments rely on heterogeneous adaptations shaped by residual vision and experience.  

\textbf{Methods} We conducted a comparative eye-tracking study of fully sighted, low vision, blind, and fully blind participants navigating outdoor routes. Using a wearable eye tracker, we quantified fixation counts, fixation rate, fixation area, direction, peak fixation location, and walking speed.  

\textbf{Results} Walking speed declined systematically with worsening vision. Fixation count increased with greater impairment, reflecting slower travel times and more frequent sampling. Fixation rate rose with worsening vision, though between-group differences were generally not significant between most groups. Fixation spatial coverage decreased along the continuum of vision loss. Fixation patterns were most consistent in the fully sighted group. Peak fixation locations were centered in fully sighted participants but shifted outward and became more variable with impairment.

\textbf{Conclusion} Gaze strategies during navigation form a graded continuum across vision groups, with fully sighted and fully blind participants at opposite poles and low vision and blind groups spanning the middle. Visual acuity alone does not predict functional gaze use, as rehabilitation experience and adaptive strategies strongly shape behavior. These findings highlight the need for personalized rehabilitation and assistive technologies, with residual gaze patterns offering insight into mobility capacity and training opportunities for safer navigation.
\end{abstract}

\section{Introduction}

Visual impairment is a major global concern, affecting nearly 300 million individuals with moderate to severe vision loss and more than 40 million with blindness worldwide \cite{visionAtlas}. Such impairments reduce independence, contribute to unemployment \cite{sherrod2014association}, increase reliance on others \cite{popescu2011age}, and diminish quality of life \cite{mckean2007severity,court2014visual}. Approximately 85\% of people with blindness or low vision (pBLV) retain some degree of residual vision, according to the American Foundation for the Blind  \cite{AFB_LowVision}. The quality and extent of this remaining vision vary, but regardless of the amount, it often plays a critical role in daily functioning. Residual vision provides a natural and immediate means of perceiving the environment, integrating with the additional perceptual systems \cite{marr2010vision,granlund2013signal} and serving as the foundation for rehabilitation programs aimed at maximizing independence and quality of life \cite{Morse2018}.

Independent mobility is among the greatest challenges faced by pBLV \cite{gallagher2011mobility, hamed2023exploratory}. Navigating city streets requires constant monitoring of surroundings, anticipation of hazards, and rapid adjustment to dynamic elements. For pBLV, these demands are magnified, and mobility difficulties are among the most frequently reported barriers to independence \cite{sherrod2014association}. A wide range of computer vision–based systems and sensory substitution devices have been developed to assist navigation, offering functions such as obstacle detection \cite{feng2025robust}, sign recognition \cite{feng2023commute}, localization \cite{ng2022real}, and wayfinding support \cite{hao2022detect}. While valuable, these tools remain limited. Sensory substitution is inherently slower than the visual channel, even when impaired, because translated signals require additional cognitive processing \cite{kristjansson2016designing, feng2025haptics}. Moreover, most systems are designed for specific tasks, whereas outdoor navigation demands the simultaneous integration of many capabilities. No single technology replicates the breadth and efficiency of the human visual system \cite{scheirer2014perceptual}. Rehabilitation services, particularly orientation and mobility training, therefore remain central. These programs emphasize structured scanning, hazard anticipation, and coordinated use of residual vision with mobility aids \cite{welsh1981foundations}. Crucially, they integrate residual vision into strategies rather than replacing it, helping individuals maximize the functional value of their vision while developing adaptive techniques for safety. The success of such training, however, depends on understanding how people with different levels of vision allocate attention and process information during real-world navigation.

Eye movements provide a powerful way to study adaptive strategies. Fixation patterns reveal what information people prioritize and how they adapt to task demands. For individuals with visual impairment, eye tracking can uncover compensatory behaviors that are not captured by broader mobility outcomes, such as walking speed. Prior studies have shown that people with low vision often increase fixation frequency and adopt specialized scanning techniques to compensate for lost fields of view  \cite{wang2023understanding}. Much of this evidence, however, comes from controlled laboratory tasks such as reading, obstacle detection, or driving simulations, where environments are simplified and predictable \cite{wang2023understanding, turano2001direction,nieboer2023eye}. A handful of outdoor studies have added important insights. Research has shown that participants with low vision often direct gaze downward or toward nearby surfaces \cite{matsuda2021gazing,freedman2019gaze}, while fully sighted individuals maintain a more forward-looking focus \cite{turano2001direction}. Other work has reported narrower scanning ranges \cite{vargas2006eye} or altered fixation allocation during tasks such as street crossing \cite{geruschat2006gaze}. These findings highlight how vision status influences gaze strategies, but most studies have examined only a single impairment type, a single task, or limited environments. Many also group all participants with low vision together, without distinguishing between different degrees of residual vision. As a result, the continuum of strategies across fully sighted, low vision, and blind individuals remain poorly understood.

To address these gaps, we examined gaze behavior during real-world, long-range navigation in New York City among fully sighted, low vision, blind and fully blind participants. Using a wearable eye tracker, we recorded fixations as participants walked a standardized urban route under natural conditions. We analyzed both mobility outcomes, such as walking speed, and fixation-based measures including fixation count, rate, direction, spatial coverage, peak fixation location, and fixation area shape similarity. We hypothesized that gaze behavior would systematically reflect the availability and quality of residual vision. Specifically, individuals with low vision were expected to adopt adaptive strategies driven by safety concerns, as evidenced by increased fixation counts, denser spatial coverage, and altered scanning patterns compared to fully sighted participants. Moreover, we anticipated a continuum in gaze behavior: participants with better residual vision would display patterns more similar to those of fully sighted individuals, while those with poorer residual vision would exhibit patterns approaching those of blind participants. By identifying these graded differences, this work provides new insight into compensatory visual strategies and informs both rehabilitation programs that train efficient use of residual vision and assistive technologies that aim to complement, rather than duplicate, human visual capabilities.

\section{Methods}

We conducted a real-world navigation study in New York City to examine gaze behavior across four functional vision groups: fully sighted, low vision, blind, and fully blind. Participants walked a standardized urban route while wearing Pupil Invisible glasses, a head-mounted eye tracker that recorded fixations continuously. Each participant completed two trials under typical summer daylight conditions, using their preferred mobility aids. Data collection focused on both mobility outcomes and fixation-based measures. The following subsections detail participant characteristics, experimental apparatus, and walking procedures.

\subsection{Participants}
Participants were recruited to evaluate gaze behavior during real-world navigation. According to the World Health Organization’s (WHO) \textit{International Classification of Diseases 11 (ICD-11)}, distance vision impairment is classified based on best-corrected visual acuity in the better-seeing eye and, in some cases, on visual field extent \cite{vaishali2020understanding, clevelandclinic_legallyblind}:  

\begin{enumerate}
    \item \textbf{No vision impairment:} visual acuity 20/40 or better, with a normal visual field.  
    \item \textbf{Mild vision impairment (Category 1):} worse than 20/40 and equal to or better than 20/70.  
    \item \textbf{Moderate vision impairment (Category 2):} worse than 20/70 and equal to or better than 20/200.  
    \item \textbf{Severe vision impairment (Category 3):} worse than 20/200 and equal to or better than 20/400.  
    \item \textbf{Blindness (Category 4):} worse than 20/400 and equal to or better than 20/1200, or counts fingers at 1 meter. 
    \item \textbf{Blindness (Category 5):} worse than 20/1200 but with residual light perception.   
    \item \textbf{Blindness (Category 6):} complete absence of light perception (no light perception, NLP).  
\end{enumerate}

For this study, these ICD-11 categories were consolidated into four functional vision groups. The \textbf{fully sighted group} included participants with best-corrected visual acuity of at least 20/40 and a normal binocular visual field \cite{ECOO2011LowVision}. Categories 1–3 (mild, moderate, and severe impairment) were combined into a single \textbf{low vision group}, reflecting both the limited sample size in this study and the fact that the distinction between low vision and blindness is clinically more fundamental than the gradations within low vision. Functionally, this group also included participants with a corresponding visual field loss to less than 20° in the better eye with best possible correction \cite{ECOO2011LowVision}. Categories 4–5 were combined into the \textbf{blind group}, encompassing participants with very poor acuity or a corresponding visual field loss to less than 10° in the better eye, but with some residual vision such as hand motion or light perception \cite{ECOO2011LowVision}. Finally, Category 6 was defined as the \textbf{fully blind group}, representing participants with complete absence of light perception (NLP), where visual field measurement is not applicable.

Notice that although fully blind participants lack visual perception and do not respond to visual stimuli, they still generate measurable eye and head orienting behaviors, reflecting preserved oculomotor control and learned orienting responses \cite{leigh1980eye}. Their inclusion is essential, as the fully blind group provides the boundary condition at the lower end of the vision continuum. With this anchor, we can capture the full spectrum of navigation-related fixation behaviors---from a fully sighted template to a non-visual oculomotor pattern---and assess where low vision and blind participants fall along that continuum. This design makes it possible to evaluate how navigation strategies shift with progressive vision loss and to clarify whether low vision participants adopt sighted-like or blind-like gaze patterns.

Exclusion criteria comprised significant cognitive dysfunction (defined as a score below 24 on Folstein’s Mini-Mental Status Examination), prior neurological illness, complex medical conditions, substantial mobility restrictions, use of walkers or wheelchairs, and pregnancy.

All procedures were reviewed and approved by the Institutional Review Board at Lighthouse Guild. Before participation, individuals received detailed information about the study objectives and procedures, after which written informed consent was obtained. Participants were then familiarized with the data collection equipment and given an overview of the designated walking route to ensure full understanding of the study process.

\subsection{Apparatus}

Eye movements were recorded using the Pupil Invisible glasses, a deep learning–based eye tracking system designed for real-world applications \cite{pupil_invisible}. The device captures binocular gaze without requiring calibration and applies slippage compensation through neural network processing, ensuring reliable tracking even during head or body movement. It provides uncalibrated gaze accuracy of approximately 4.6° and outputs 2D gaze points in scene camera coordinates at 120 Hz in real time \cite{pupil_invisible, tonsen2020high}. The glasses weigh only 46.9 grams, making them comfortable for extended use in outdoor environments \cite{pupil_invisible}.

The Pupil Invisible was selected for this study because of its practical advantages for both participants and researchers. Its calibration-free design avoids the need for lengthy setup procedures, which is especially important outdoors where recalibration may otherwise be necessary if the device is disturbed. This feature is particularly beneficial for low-vision participants, where safety concerns can occasionally interrupt trials.

In addition, the device’s lightweight, glasses-like design closely resembles ordinary eyewear. This natural form factor reduces participant awareness of the equipment and allows them to behave as they normally would, including turning the head and scanning the environment freely. By combining sufficient accuracy with ecological realism, the Pupil Invisible provides an optimal balance of comfort, reliability, and validity for studying gaze behavior during real-world navigation.

\subsection{Procedures}

All participants were equipped with the eye-tracking device and instructed on how to wear it before walking along a designated route from 250 W 64th St, New York, NY 10023 (40.7744525, -73.9883542), to the subway entrance at 66 St–Lincoln Center (40.7738218, -73.9825589). The route spanned approximately 0.5 miles one way and is nearly flat, with a maximum elevation difference of 23 feet. Each one-way journey constituted one trial, and participants completed a round trip, amounting to two trials and a total walking distance of about 1 mile. Participants were permitted to use their preferred mobility aids, such as a white cane, a guide dog, or none, and to rely on their usual navigation strategies, such as listening for traffic or mental mapping. Experimenters intervened only to prevent immediate dangers, including stepping into traffic, colliding with obstacles, or risking a fall.

\begin{figure}

    \centering
    
    \includegraphics[width=14cm]{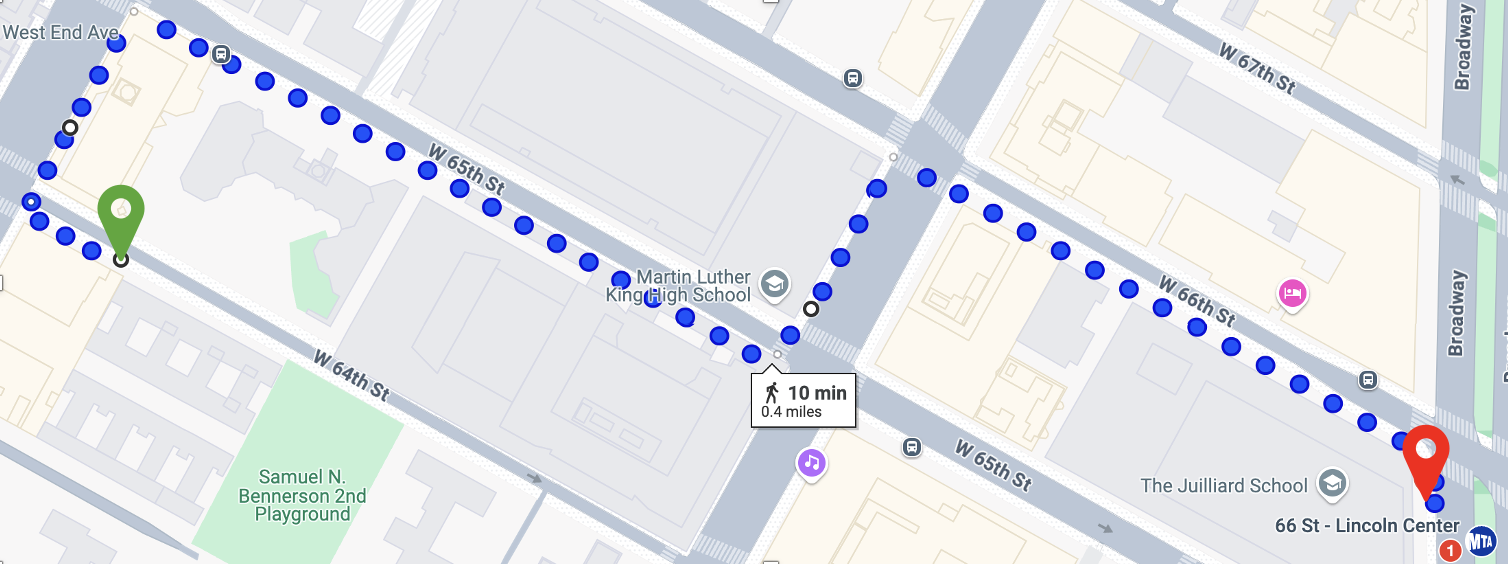}
    
    \caption{The experimental route. The green marker on the left represents the location of Lighthouse Guild, while the red marker on the right indicates the entrance to 66th St–Lincoln Center. Traveling from the green marker to the red marker constitutes Trial 1, and the reverse journey constitutes Trial 2.
    \label{figure_route}
    }
\end{figure}

The route, illustrated in Figure \ref{figure_route}, required three direction changes at crosswalks. At these points, participants were allowed to request directions or ask about the traffic light status. Experimenters also corrected participants if they began heading in the wrong direction for more than 3-4 steps. The sidewalks along the route were constructed of unpigmented concrete, the most common sidewalk material in New York City \cite{nyc_street_design_guide_2020}. These sidewalks lacked tactile clues, reflecting typical pedestrian pathways in the city. This route represents a standard last-mile commute, illustrating the distance from a transportation hub to a final destination.

The route featured a variety of static obstacles, including scaffolding zones, trees, tree guards, bike racks, bus stop shelters, fire hydrants, sign poles, light poles, and trash cans, as well as other types of street furniture and hazards. These obstacles remained in fixed locations throughout all participant trials. In addition, dynamic obstacles such as moving vehicles, pedestrians, traffic cones, and trash piles varied in both location and number during each trial. Semi-static obstacles, including parked cars, parked bikes, and food trucks, were generally in similar locations but not in exactly the same spots each day.

All trials took place between 10 a.m. and 5 p.m. on typical weekdays under sunny or cloudy weather conditions, from May to August. No water puddles were observed on the sidewalks during the trials. The daylight, temperature, and humidity were characteristic of a typical New York City summer, and no trials were conducted during extreme weather events. This setup ensured a controlled yet realistic environment for all participants.

Participants were instructed to walk along the route in their usual manner.

\subsection{Heatmap Generation}

For each recording, the Pupil Invisible system produced a fixation file containing gaze events detected in the data stream. Each record included the following fields: fixation IDs, start and end timestamps, fixation duration, fixation coordinates ($x$, $y$) in scene camera pixels, and gaze ray azimuth and elevation in degrees \cite{pupil_labs_data_format}. These outputs provided the raw data necessary to quantify fixation patterns during navigation.  

To standardize the spatial distribution of fixations across participants, fixation heatmaps were generated from these files. Heatmaps captured the frequency of fixations within discretized regions of the visual scene, serving as an intermediate representation for subsequent analyses such as coverage area, Fourier similarity, and centroid dispersion.  

Fixation coordinates $(x, y)$ were extracted from the Pupil Cloud output and mapped onto a unified $1088 \times 1080$ pixel coordinate system, corresponding to the resolution of the scene camera. This ensured that gaze data from all participants were aligned to a consistent spatial reference frame.  

All fixation points from a given trial were then aggregated into a two-dimensional histogram of gaze distribution. The coordinate space was divided into a $50 \times 50$ grid, with each bin recording the number of fixations falling within its boundaries. The resulting histogram provided a raw heatmap representing fixation density across the visual scene.  

To reduce pixel-level noise and produce smoother fixation density estimates, the raw heatmap was convolved with a Gaussian kernel ($\sigma = 1$). Gaussian smoothing replaced each bin’s value with a weighted average of its neighbors, with weights determined by a Gaussian function centered on the bin. This step ensured that the heatmap reflected continuous fixation fields rather than sparse, discretized points, improving robustness to small variations in fixation placement.  

To account for individual differences in fixation count, heatmaps were normalized by dividing each bin value by the maximum bin count for that trial:  
\[
\text{Normalized Heatmap Value}_{i,j} = \frac{\text{Bin Count}_{i,j}}{\max(\text{Bin Count})}.
\]
This normalization rescaled each heatmap to the range $[0, 1]$, with 1 representing the region of highest fixation density. Normalization was necessary to ensure that group-level comparisons reflected the \textit{relative spatial distribution of visual attention}, rather than being confounded by absolute fixation counts.  

The resulting normalized and smoothed heatmaps provided a standardized representation of gaze allocation and formed the basis for all subsequent analyses described in Section~\ref{sec:data_analysis}.

\subsection{Data Analysis}
\label{sec:data_analysis}
Building on the normalized fixation heatmaps and other eye-tracking measures, several quantitative analyses were conducted to evaluate navigation behavior across groups. Group-level differences across vision groups were assessed using Welch’s one-way ANOVA, which does not assume equal variances. Pairwise contrasts were evaluated with Games–Howell post-hoc tests, and Brunner–Munzel nonparametric tests were additionally performed to confirm robustness under non-normal distributions. This statistical framework was applied consistently across all metrics.

\subsubsection{Data exclusion}

Participant data were excluded from analysis if more than 50\% of the trip was missing due to device failure or experiment termination. In the experiments, missing data did occur, but all exclusions were due solely to technical problems with the recording devices rather than participant-related factors. For instance, the scene camera, magnetically attached to the arms of the eyeglasses, occasionally became disconnected when participants adjusted their glasses by pulling on the arm with the mounted scene camera. Such adjustments caused temporary disconnections, resulting in missing video frames. Additionally, some participants were unaware of the cable's placement, and their body movements occasionally pulled the cable, leading to unintentional disconnections. Once disconnected, the recording paused, causing a complete data loss for the remainder of the trip.

These issues were identified during the early phases of the experiment, prompting the experimenter to take proactive measures. Participants were instructed to avoid adjusting the glasses by pulling on the arm with the scene camera, and the cable connection was closely monitored. These issues were concentrated in the early portion of data collection, which included both fully sighted and low vision participants. In this first half of the study, several participants experienced complete data loss due to device disconnections. After adjustments to procedures (e.g., instructing participants not to adjust the glasses by the scene camera arm and closer cable monitoring), no further data loss occurred in the later half of recruitment.

To ensure the integrity of the analysis, exclusions were based solely on objective criteria related to data completeness and were not influenced by the content or outcomes of the recordings. These exclusions were necessary to maintain the quality and reliability of the dataset, and we are confident they did not introduce bias into the study's results. Participants with unverified or incomplete medical records confirming their vision status were also excluded from the analysis.

\subsubsection{Walking Speed Analysis}

The first metric examined was walking speed, computed as an objective measure of mobility performance during the navigation task. Walking speed was selected because it is a simple yet robust indicator of functional mobility and navigation efficiency in real-world environments and has been widely used in vision and rehabilitation research to quantify mobility outcomes.

For each participant, walking speed was calculated as the ratio of total route distance to travel time, expressed in meters per second (m/s). Travel time was defined as the duration from the starting point to the destination, or vice versa. Since all participants followed the same route, the travel distance was identical across trials.

To assess group-level differences, mean walking speeds were compared across all four groups. Group-level differences were tested using the statistical framework described above.

\subsubsection{Fixation Counts Analysis}

Fixation count was included as a key metric to capture the extent of visual sampling during navigation. The total number of fixations made by each participant provides a direct measure of how frequently the visual system pauses to extract information from the environment. Higher fixation counts may reflect greater cognitive or perceptual effort, particularly in individuals with limited or no residual vision, while lower counts may indicate more efficient information uptake.

For each participant, fixation count was calculated as the total number of fixations detected during a single one-way trial. Each participant completed two trials, and fixation counts were computed separately for each trial. Fixation counts were derived directly from Pupil Cloud’s fixation output by subtracting the fixation ID at the start point from the fixation ID at the endpoint, since the software assigns fixation IDs in ascending numerical order for each detected fixation.

To evaluate group-level differences, fixation counts were compared across four groups. Group-level differences were tested using the statistical framework described above.

\subsubsection{Fixation Rate Analysis}

Fixation rate was included as a temporal measure of visual behavior, defined as the number of fixations per minute during navigation. This metric complements fixation counts  by capturing how frequently participants sampled their visual environment relative to time, independent of trial duration.  

For each participant, fixation rate was calculated by dividing the total number of fixations detected in a trial by the corresponding trial duration (in minutes). Rates were computed separately for each trial and then averaged within groups to allow for between-group comparisons.  

To evaluate within-trial variability, we divided each trial into 60-second bins, computed the fixation rate within each bin, and then calculated the difference between the maximum and minimum rates across the trip. This max–min difference captured the range of sampling rates expressed during a single trial and is referred to as fixation rate fluctuation in the later sections.

Group-level analyses were performed across all four groups. Group-level differences were tested using the statistical framework described above.

\subsubsection{Fixation Spatial Coverage Analysis}
To quantify the extent of visual exploration within the scene, we measured the spatial coverage of fixation distributions. Spatial coverage was defined as the proportion of the scene with fixation density $\geq 0.001$ of each participant’s maximum, which excluded low-intensity noise while retaining meaningful coverage. Because each trial contained on the order of 1000 fixations, this relative threshold approximately corresponded to at least one fixation per bin, thereby ensuring that only regions with genuine gaze allocation were included while filtering out bins driven by noise or incidental fixations.

As described in Section~3.2, fixation heatmaps were generated by mapping gaze points onto a $1088 \times 1080$ pixel coordinate system representing the scene camera’s resolution. This coordinate space was discretized into a $50 \times 50$ grid, where each grid cell (approximately $21.76 \times 21.6$ pixels) represented a local region of the scene. Fixations falling within each grid cell were accumulated to produce a two-dimensional histogram of gaze density.  

The raw heatmap was normalized by dividing each bin count by the maximum bin count in that trial, resulting in values between 0 and 1. A binary mask was then constructed by marking all grid cells with normalized values greater than or equal to 0.001. Spatial coverage was computed as the percentage of grid cells in this mask relative to the total $50 \times 50 = 2500$ cells.  

Group-level comparisons of fixation spatial coverage were conducted across all four groups. Group-level differences were tested using the statistical framework described above.

\subsubsection{Fixation Direction Analysis}

In addition to fixation counts, we analyzed the spatial orientation of gaze by quantifying fixation azimuth and fixation elevation. These metrics capture the horizontal and vertical distribution of visual attention during navigation and provide insight into how participants allocate gaze relative to the forward-facing direction of travel. Examining these values helps to identify whether different vision groups adopt systematic gaze strategies, such as preferentially directing attention toward the ground, horizon, or lateral features of the environment.

Fixation azimuth represents the horizontal angle (in degrees) of the gaze ray relative to the center of the scene camera’s field of view. Positive azimuth values indicate a gaze direction to the right of center, whereas negative values indicate a gaze direction to the left. Fixation elevation refers to the vertical angle (in degrees) of the gaze ray relative to the center of the scene camera’s field of view. Positive elevation values correspond to gaze directions above the center, while negative values correspond to gaze directions below. Both azimuth and elevation values were derived from the Pupil Cloud data output for each detected fixation, following the manufacturer’s data format specifications \cite{pupil_labs_data_format}.

For each participant, mean fixation azimuth and mean fixation elevation were computed across the duration of a trial. These trial-level averages were then aggregated across participants within each vision group. Statistical analyses were conducted to test for group-level differences. Group-level differences were tested using the statistical framework described above.

\subsubsection{Peak Fixation Location Analysis}

To examine spatial biases in visual attention, we analyzed the location of peak fixation density relative to the center of the scene camera’s field of view. For each participant, a fixation heatmap was first generated and normalized. The bin with the maximum normalized value was identified as the peak fixation location, corresponding to the region of greatest visual concentration during navigation. Because each trip contained more than 1000 fixations, we restricted the analysis to bins with fixation density values above 0.001, ensuring that identified peaks reflected at least one fixation event.

The scene camera output was divided into a $50 \times 50$ grid, with each bin corresponding to approximately $21.76 \times 21.6$ pixels. This discretization provided sufficient spatial resolution while reducing noise from frame-to-frame variability. The centroid of the peak bin was extracted as the participant’s peak fixation location.

Peak fixation locations were plotted for each group and visualized with convex hull boundaries to characterize the spread of fixation hotspots within groups. Distances from each participant’s peak fixation location to the geometric center of the frame were computed to quantify spatial displacement. Group-level comparisons of these distances were conducted. Group-level differences were tested using the statistical framework described above.

\subsubsection{Fixation Area Shape Similarity Analysis}

To evaluate whether participants with different visual status exhibited consistent spatial patterns of visual exploration, we analyzed the shape similarity of fixation coverage areas across groups. Shape similarity was quantified using Fourier descriptor–based metrics, which provide a robust representation of a shape’s contour in the frequency domain \cite{cortese1996perceptual}. This approach ensures invariance to translation, rotation, and scale, thereby enabling direct comparisons across participants.

For each trial, a fixation heatmap was generated on a $50 \times 50$ grid and smoothed with a Gaussian kernel ($\sigma=1$). The heatmap was normalized, and a binary mask was obtained by thresholding at $0.1\%$ of the participant’s maximum fixation density, ensuring that regions included in the mask reflected meaningful fixation clusters. This parameterization (grid resolution, Gaussian smoothing, and threshold level) matched the visualization settings used for heatmaps. Since most trials contained at least 1000 fixations, a threshold of $0.1\%$ corresponded to at least one fixation in a given bin. The boundary contour of the largest connected region was then extracted and uniformly resampled to 1024 points. The resampled contour was converted into a complex-valued sequence and transformed into the frequency domain using the discrete Fourier transform (DFT). The first 20 Fourier coefficients (excluding the DC component) were retained, capturing the coarse structural features of the fixation contour while discarding high-frequency noise, consistent with prior applications of Fourier descriptors in shape analysis. Their magnitudes were normalized to eliminate sensitivity to starting point, scaling, and rotation.

Pairwise similarity between two fixation area shapes was computed using an inverse-distance metric:  
\[
S = \frac{1}{1 + \left\| D_1 - D_2 \right\|},
\]
where \(D_1\) and \(D_2\) are the Fourier descriptor vectors for the two shapes. Similarity values closer to 1 indicate greater overlap in shape structure, whereas values closer to 0 indicate dissimilarity.

Group-level differences in similarity distributions were statistically evaluated. Group-level differences were tested using the statistical framework described above.

\subsubsection{Individual-Level Fixation Analysis}

Beyond group-level comparisons, we examined how navigation outcomes varied with individual differences in visual acuity. Snellen OD and OS values were converted to decimal acuity (for example, 20/40 = 0.5 and 20/200 = 0.1). The better eye represented functional vision. Decimal acuity was then transformed to the logarithm of the minimum angle of resolution (logMAR) scale using
\[
\text{logMAR} = -\log_{10}(\text{decimal acuity}).
\]
On this scale, 0.0 corresponds to normal vision (20/20), and higher values indicate worse acuity. For example, 20/200 maps to logMAR = 1.0 and 20/400 maps to logMAR = 1.3. To incorporate categorical cases, we assigned fixed anchors that are not derived from Snellen fractions: hand motion (HM) = 2.0, light perception (LP) = 2.3, and no light perception (NLP) = 3.0. Fully sighted participants therefore scored 0.0, while those with little or no residual vision mapped to higher logMAR values.

This approach preserves the clinical convention where normal vision is 0.0 and worse vision increases monotonically. Individual outcomes, including walking speed, fixation count, fixation rate, and fixation spatial coverage, were plotted against each participant’s logMAR score. Identical values, such as the cluster of sighted participants at 0.0, were jittered slightly to reduce overlap and labeled once for clarity. This representation highlights a continuum across the spectrum of vision loss: blind participants clustered at the high logMAR end, sighted participants at the lower bound, and those with partial vision distributed between these extremes according to their measured acuity.

\section{Results}

\subsection{Participant Demographics}
We recruited 24 participants with blindness or low vision and 18 fully sighted participants. The 24 participants with blindness or low vision were classified into the low vision, blind, and fully blind groups. Among them, 5 were excluded due to substantial data loss from eye-tracking device failures, leaving 19 participants (10 low vision, 4 blind, and 5 fully blind). Most of these 19 participants reported having received orientation and mobility training (18/19), training in the use of low vision devices (15/19), training in daily living skills (12/19), and vocational rehabilitation (12/19). All participants with blindness or low vision reported at least one form of rehabilitation training and described themselves as physically active. Nearly all (18/19) traveled independently to the testing site without accompanying aids. Within this group, 2 participants used guide dogs, 2 reported not using any mobility aid, and 15 used white canes. Of the 18 fully sighted participants, eye-tracking data were missing for 6 due to device failure, leaving 12 participants with usable data, all of whom were included in the analysis.

Note that one participant (Participant ID LV2) with an eye disease had a best-eye visual acuity of 20/33, but the visual field was restricted, so this participant was classified as low vision in the this study.

Details are summarized in Table~\ref{tab:participant_summary_updated}.

\begin{table*}[t]
\centering
\resizebox{\textwidth}{!}{%
\begin{tabular}{|l|l|c|c|l|c|c|l|l|}
\hline
\textbf{Condition} & \makecell{\textbf{Participant} \\ \textbf{ID}} & \textbf{Age} & \makecell{\textbf{Years with} \\ \textbf{Vision Problem}} & \textbf{Diagnosis} & \textbf{Visual Acuity (Right)} & \textbf{Visual Acuity (Left)} & \textbf{Field} & \textbf{Mobility Aid} \\ \hline

\multirow{12}{*}{Fully Sighted} 
& FS1 & 34 & 0  & Normal  & 20/20* & 20/20* & Full & None \\ \cline{2-9}  
& FS2 & 24 & 0  & Normal  & 20/20* & 20/20* & Full & None \\ \cline{2-9}  
& FS3 & 23 & 0  & Normal  & 20/20* & 20/20* & Full & None \\ \cline{2-9}  
& FS4 & 73 & 0  & Normal  & 20/20* & 20/20* & Full & None \\ \cline{2-9}  
& FS5 & 29 & 0  & Normal  & 20/20* & 20/20* & Full & None \\ \cline{2-9}  
& FS6 & 60 & 0  & Normal  & 20/20* & 20/20* & Full & None \\ \cline{2-9}
& FS7 & 54 & 0  & Normal  & 20/20* & 20/20* & Full & None \\ \cline{2-9} 
& FS8 & 69 & 0  & Normal  & 20/20* & 20/20* & Full & None \\ \cline{2-9} 
& FS9 & 33 & 0  & Normal  & 20/20* & 20/20* & Full & None \\ \cline{2-9} 
& FS10 & 23 & 0  & Normal  & 20/20* & 20/20* & Full & None \\ \cline{2-9} 
& FS11 & 56 & 0  & Normal  & 20/20* & 20/20* & Full & None \\ \cline{2-9} 
& FS12 & 20 & 0  & Normal  & 20/20* & 20/20* & Full & None \\ \hline

\multirow{11}{*}{Low Vision} 
& LV1 & 51 & 6 & Retinal detachment with multiple breaks, bilateral & 20/200 & 20/250 & Central loss & White cane \\ \cline{2-9}
& LV2 & 27 & 21 & Stargardt’s Disease & 20/300 & 20/33 & Central scotoma & White cane \\ \cline{2-9}
& LV3 & 68 & 5 & Cone Dystrophy & 20/50 & 20/400 & Central scotoma & White cane \\ \cline{2-9}
& LV4 & 69 & 69 & Macular Degeneration & 20/250 & 20/250 & Central scotoma & White cane \\ \cline{2-9}
& LV5 & 70 & 9 & Age-related Macular Degeneration & 20/250 & 20/250 & Central scotoma & White cane \\ \cline{2-9}
& LV6 & 29 & 28 & Juvenile X-linked Retinoschisis & 20/200 & 20/200 & Central loss & None \\ \cline{2-9}
& LV7 & 53 & 20 & Con-rod Dystrophy & 20/250 & 20/400 & Central loss & White cane \\ \cline{2-9}
& LV8 & 59 & 59 & Glaucoma, Left Retinal Detachment & 20/200 & No light perception & Peripheral loss & White cane \\ \cline{2-9}
& LV9 & 42 & 42 & Albinism, Nystagmus & 20/400 & 20/400 & Moderate restriction & None \\ \cline{2-9}
& LV10 & 53 & 53 & Congenital Nystagmus & 20/250 & 20/250 & Severe restriction & White cane \\ \hline

\multirow{4}{*}{Blind} 
& BL1 & 62 & 35 & Retinitis Pigmentosa & Hand motion & Light perception & Severe constriction & White cane \\ \cline{2-9}
& BL2 & 73 & 73 & Glaucoma & 20/500 & Light perception & Peripheral loss & White cane \\ \cline{2-9}

& BL3 & 61 & 61 & Acquired hypothyroidism; Macular atrophy, retinal  & Light perception & Light perception & None & White cane \\ \cline{2-9}

& BL4 & 64 & 58 & Leber congenital amaurosis & Light perception & Light perception & None & White cane \\ \hline

\multirow{5}{*}{Fully Blind} 
& FB1 & 66 & 66 & Acute congenital glaucoma & No light perception & No light perception & None & White cane \\ \cline{2-9}
& FB2 & 70 & 68 & Retinopathy of prematurity & No light perception & No light perception & None & Guide dog \\ \cline{2-9}
& FB3 & 71 & 61 & Retinopathy of prematurity & No light perception & No light perception & None & White cane \\ \cline{2-9}
& FB4 & 82 & 82 & Uveitis, cataract & No light perception & No light perception & None & Guide dog \\ \cline{2-9}
& FB5 & 57 & 52 & Retina detachment, glaucoma, and cataracts & No light perception & No light perception & None & White cane \\ \hline

\end{tabular}%
}
\caption{Participant details grouped by vision condition, with  classification into four groups: fully sighted, low vision, blind, and fully blind.}
\label{tab:participant_summary_updated}
\end{table*}

\subsection{Heatmaps}
\begin{figure}
    \centering
    \includegraphics[width=14cm]{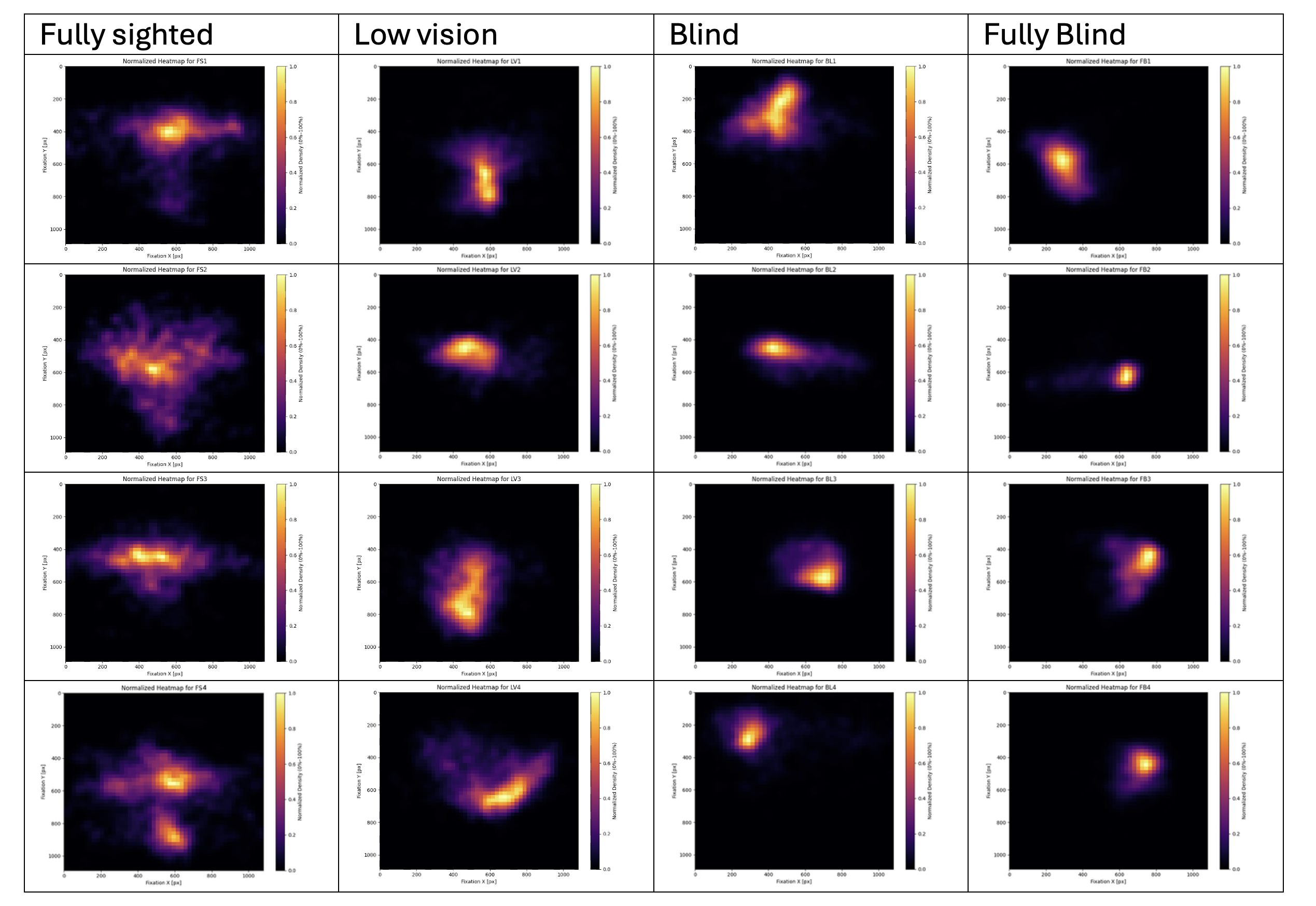}
    \caption{Representative Heatmaps of fixations for participants in the fully sighted, low vision, blind and fully blind groups. Each heatmap represents a trial from an individual participant.}
    \label{figure_heat_map_all}
\end{figure}

Figure \ref{figure_heat_map_all} shows representative heatmaps of fixation distributions, displaying the first five trials from each group in ascending order of participant ID. We limited the figure to the first four participants per group to maintain visual clarity, since including too many heatmaps in a single figure would make each one too small to interpret. Among fully sighted participants, the heatmaps consistently reveal a T-shaped distribution, reflecting a systematic pattern of allocating gaze between the near ground and distant horizon. In contrast, the heatmaps of low vision participants exhibit considerable variability, with no single dominant pattern emerging across individuals. For blind and fully blind participants, fixation coverage was minimal and confined to a small, rounded area, reflecting the highly restricted nature of their visual sampling.

\subsection{Walking Speed}

\begin{figure}[h]
    \centering
    \includegraphics[width=13cm]{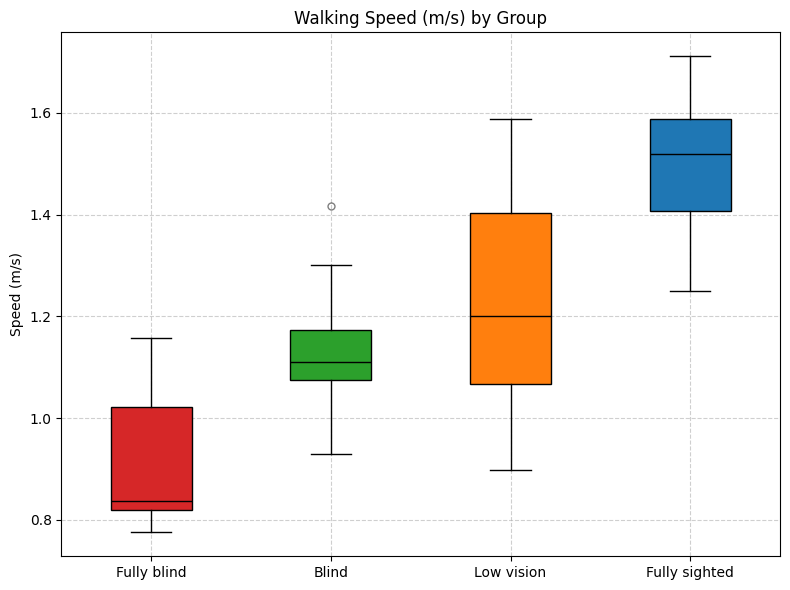}
    \caption{Box plot of walking speed (m/s) across groups. }
    \label{walking_speed_boxplot}
\end{figure}

Walking speed differed systematically across the four vision groups (Figure~\ref{walking_speed_boxplot}). 
The fully sighted group walked the fastest ($M = 1.50$ m/s, variance = 0.01), followed by the low vision group ($M = 1.23$ m/s, variance = 0.05), the blind group ($M = 1.14$ m/s, variance = 0.02), and the fully blind group ($M = 0.91$ m/s, variance = 0.02).  

Games--Howell pairwise comparisons confirmed significant differences for most contrasts. 
Fully sighted participants walked significantly faster than fully blind  ($\Delta M = 0.59$, $p < 0.001$, Hedges’ $g = -4.57$), blind ($\Delta M = 0.36$, $p < 0.001$, $g = 2.71$), and low vision participants ($\Delta M = 0.27$, $p < 0.001$, $g = 1.57$). Fully blind participants also walked significantly slower than blind  ($\Delta M = -0.23$, $p = 0.03$, $g = -1.51$) and low vision participants ($\Delta M = -0.32$, $p < 0.001$, $g = -1.54$). In contrast, no significant difference was observed between blind and low vision participants ($\Delta M = -0.09$, $p = 0.65$, $g = -0.41$).   

Brunner--Munzel tests yielded consistent results, confirming significant differences between fully blind and blind ($BM = 3.74$, $p = 0.004$), fully blind and low vision ($BM = 5.97$, $p < 0.001$), blind and fully sighted ($BM = 11.80$, $p < 0.001$), and low vision and fully sighted ($BM = 5.59$, $p < 0.001$). No significant difference was observed between blind and low vision participants ($BM = 0.92$, $p = 0.37$).

Together, these findings indicate a graded decline in walking speed with increasing severity of vision loss, with the most pronounced slowing observed in the fully blind group.

\subsubsection{Individual-level Walking Speed}

\begin{figure}[h]
    \centering
    \includegraphics[width=18cm]{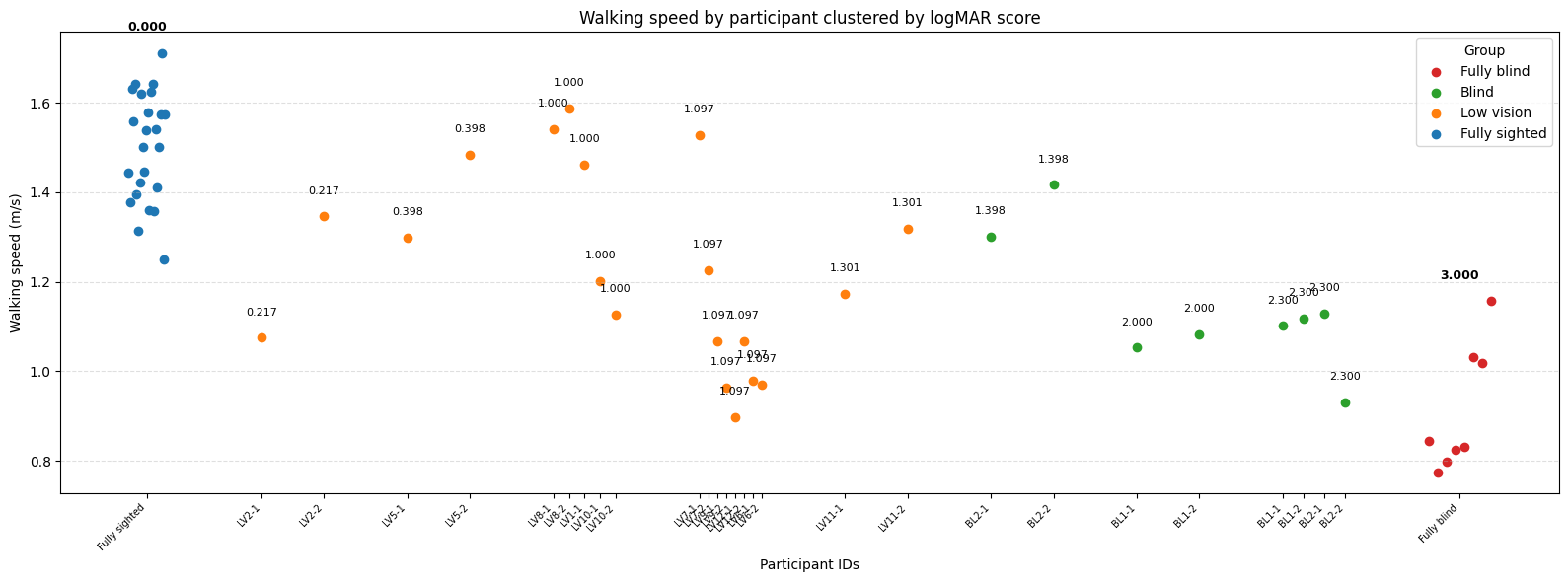}
    \caption{Participant-level walking speed. Each dot represents one participant and trial, grouped by vision category. Numbers next to dots indicate visual acuity scores for that participant.}
    \label{walking_speed_individuals}
\end{figure}

Inspection of individual-level distributions (Figure~\ref{walking_speed_individuals}) reinforces these group-level trends. Fully sighted participants consistently walked faster, with no trial below 1.2 m/s. In contrast, fully blind participants walked the slowest, with even the fastest individual remaining below 1.2 m/s.  

Among low vision participants, visual acuity appeared to influence performance. A visual acuity score of 1.0 emerged as a cutoff: equal or below this score, the majority of trials (77\%) were faster than 1.2 m/s, whereas above this score, most trials (70\%) were slower than 1.2 m/s.  

For the blind group, the participant with the best visual acuity (logMAR score of 1.389) walked faster than all others in the group and was the only individual to exceed 1.2 m/s. 

\subsection{Fixation Counts}

Fully blind participants exhibited the highest mean fixation count ($M = 2613.00$, variance = 692{,}091.00), followed by blind participants ($M = 2148.75$, variance = 108{,}662.79), low vision participants ($M = 1614.53$, variance = 187{,}865.93), and fully sighted participants ($M = 1451.83$, variance = 48{,}379.19).

Games--Howell comparisons indicated significant differences between fully blind and fully sighted ($\Delta M = 1161.17$, $p = 0.01$, Hedges’ $g = 2.45$), fully blind and low vision ($\Delta M = 998.47$, $p = 0.03$, $g = 1.66$), blind and fully sighted ($\Delta M = -696.92$, $p < 0.001$, $g = -2.72$), and blind and low vision ($\Delta M = 534.22$, $p = 0.01$, $g = 1.27$). In contrast, no significant differences were observed between fully blind and blind ($\Delta M = 464.25$, $p = 0.45$, $g = 0.68$) or between fully sighted and low vision ($\Delta M = -162.69$, $p = 0.46$, $g = -0.48$).  

Brunner--Munzel tests yielded consistent results, confirming significant differences between fully blind and fully sighted ($BM = -7.99$, $p < 0.001$), fully blind and low vision ($BM = -4.03$, $p = 0.002$), blind and fully sighted ($BM = 16.11$, $p < 0.001$), and blind and low vision ($BM = -4.43$, $p = 0.002$). No significant differences were observed between fully blind and blind ($BM = -1.13$, $p = 0.29$) or between fully sighted and low vision ($BM = 1.21$, $p = 0.24$).

\begin{figure}[h]
    \centering
    \includegraphics[width=13cm]{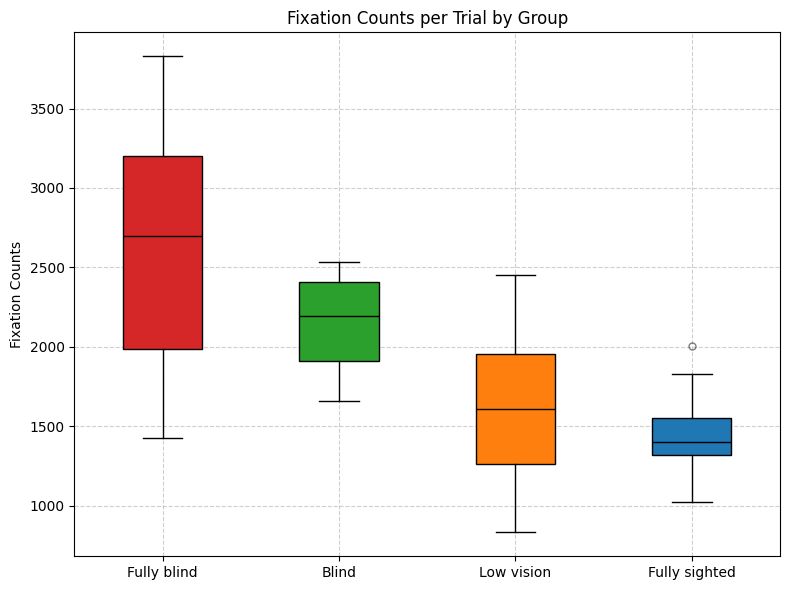}
    \caption{Box plot showing fixation count distribution across groups. Fully blind group exhibited the highest mean counts, followed by blind, low vision, then fully sighted.}
    \label{fixation_count_box_plot}
\end{figure}

As shown in Figure~\ref{fixation_count_box_plot}, the distributions of fixation counts were clearly separated across groups.

\subsubsection{Individual-level Fixation Count}

\begin{figure}[h]
\centering
\includegraphics[width=18cm]{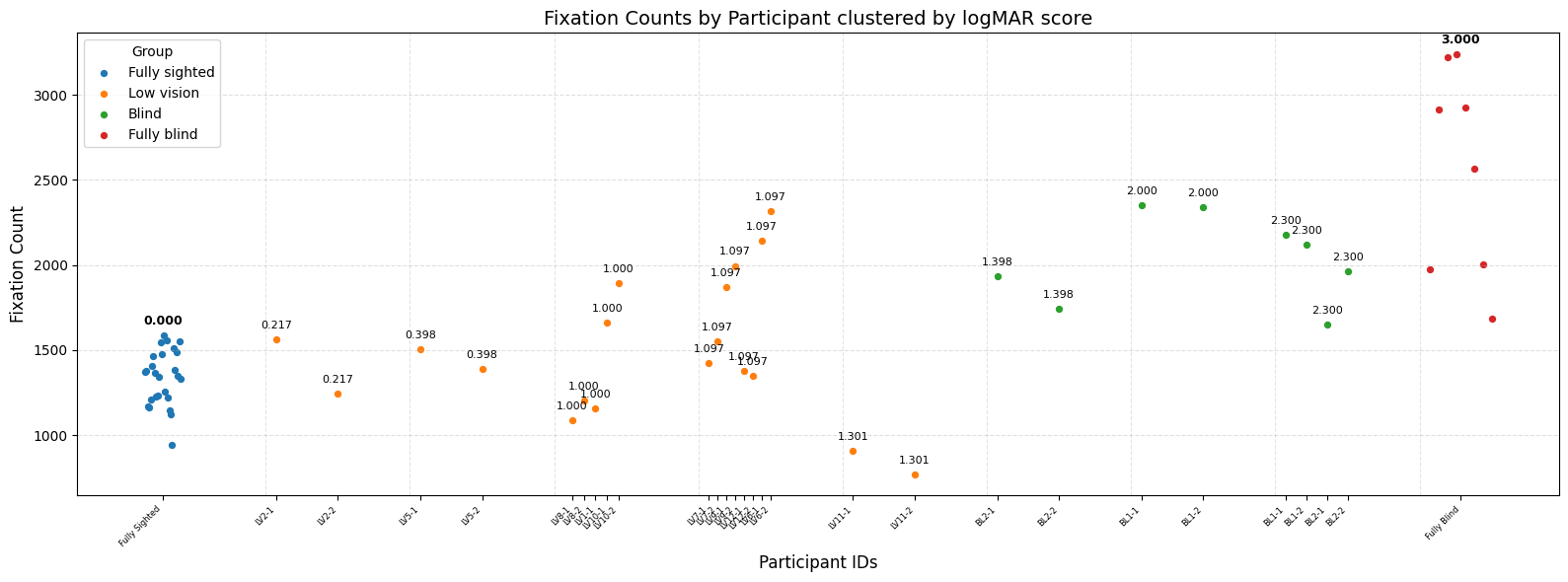}
\caption{Fixation counts per participant, grouped by vision category. Each dot represents one trial. Clusters with similar fixation counts are compactly labeled, while higher values are annotated individually.}
\label{fixation_count_dot_plot}
\end{figure}

Inspection of individual-level distributions (Figure~\ref{fixation_count_dot_plot}) shows that fixation counts in the fully sighted group clustered tightly around approximately 1452 fixations per trial, reflecting relatively consistent sampling across participants. In contrast, the low vision group exhibited much greater variability, with widely scattered values across individuals. The blind group showed fixation counts similar to the lower range of the low vision group, indicating comparatively reduced sampling. The fully blind group had the highest fixation counts overall.

\subsection{Fixation Rate}

\begin{figure}[h]
    \centering
    \includegraphics[width=13cm]{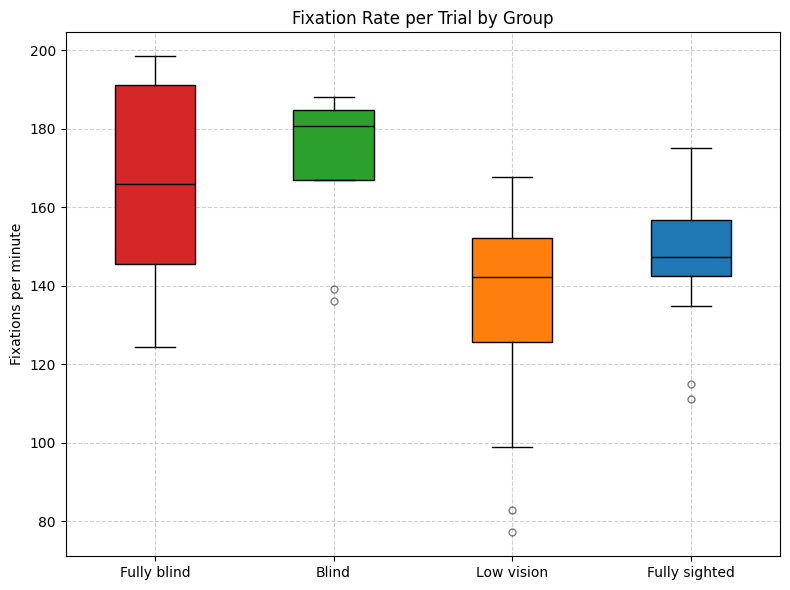}
    \caption{Box plot of fixation rates (fixations per minute) across groups.}
    \label{fixation_rate_box_plot}

\end{figure}

The mean fixation rate was highest in the blind group ($M=171.38$, variance = 450.35), followed by the fully blind group ($M=164.88$, variance = 783.96), the fully sighted group ($M=148.33$, variance = 235.28), and the low vision group ($M=132.67$, variance = 699.63). As illustrated in Figure~\ref{fixation_rate_box_plot}, the distributions showed partial overlap across groups but also indicated group-level trends.

Games--Howell pairwise comparisons identified a significant difference between blind and low vision participants ($\Delta M = 38.71$, $p = 0.01$, Hedges' $g = 1.50$). All other pairwise contrasts, including fully blind versus low vision ($p = 0.06$), were non-significant ($p > 0.07$).

Brunner--Munzel tests yielded consistent findings, confirming significant differences between blind and low vision ($BM = -3.77$, $p = 0.01$) and between fully blind and low vision ($BM = -2.73$, $p = 0.02$). Other contrasts did not reach significance.

Together, these results suggest that fixation rate was generally comparable across groups, with the exception that participants with low vision exhibited a lower sampling rate relative to blind participants.

\subsubsection{Fixation Rate Fluctuation}

Group means of this fluctuation were as follows: Blind = 44.7 fix/min, Fully sighted = 49.8 fix/min, and Low vision = 61.0 fix/min. Variances diverged more clearly (Blind = 156.8; Fully sighted = 483.3; Low vision = 264.7). Blind participants showed the most constrained dynamics, with consistently low variability in fixation rate. Fully sighted participants exhibited the greatest variability, consistent with flexible adjustment of fixation behavior to environmental complexity. Low vision participants fell between these extremes, reflecting heterogeneous strategies shaped by the nature of residual vision. Note that the fully blind group showed a mean fluctuation of 114.25 fix/min (variance = 1519.07). Because of the small sample size and the fact that their fixation activity reflects oculomotor behavior without visual input, this variability should not be considered stimulus-driven.

\subsubsection{Individual-level Fixation Rate}

\begin{figure}[h]
    \centering
    \includegraphics[width=18cm]{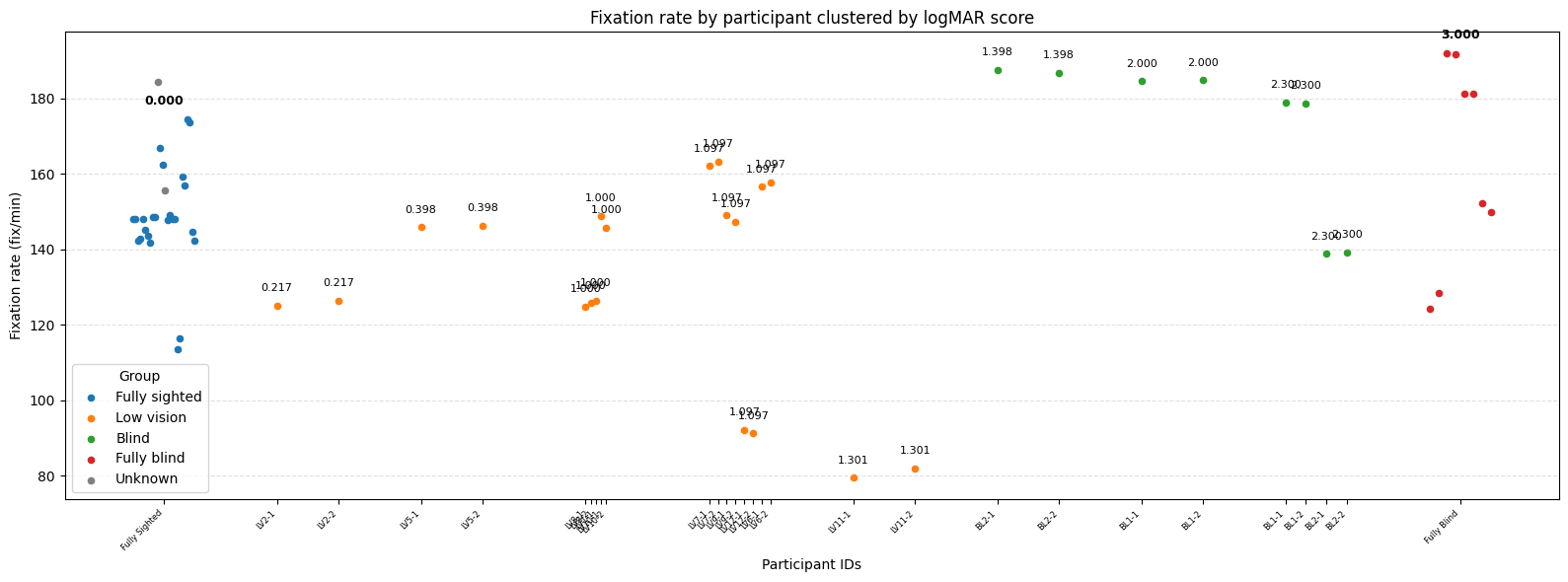}
    \caption{Participant-level fixation rate. Each dot represents one participant and trial, grouped by vision category. Numbers next to dots indicate visual acuity scores for that participant.}
    \label{fixation_rate_individuals}
\end{figure}

Inspection of participant-level distributions (Figure~\ref{fixation_rate_individuals}) reinforces this pattern. Across all groups, participants maintained broadly similar fixation rates, although the low vision participants exhibited greater variability. In contrast, blind and fully blind participants demonstrated markedly higher fixation rates than the low vision group.   

These findings suggest a trend in which fixation rate increases as visual function worsens.

\subsection{Fixation Spatial Coverage}

\begin{figure}[h]
    \centering
    \includegraphics[width=14cm]{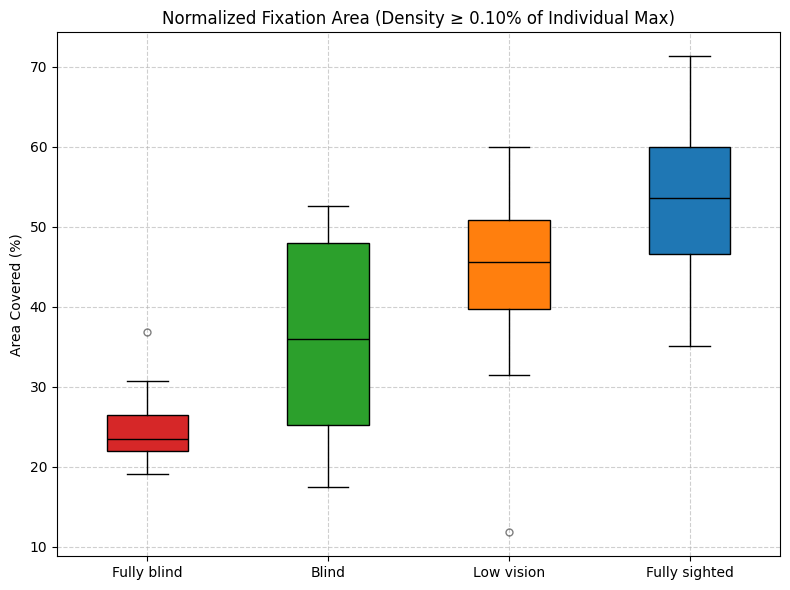}
    \caption{Box plot of normalized fixation area coverage across groups.}
    \label{fixation_area}
\end{figure}

Normalized area coverage was highest for fully sighted participants ($M = 53.42\%$, variance = 84.60), followed by participants with low vision ($M = 44.71\%$, variance = 123.25), blind participants ($M = 36.10\%$, variance = 186.29), and fully blind participants ($M = 25.21\%$, variance = 31.23). As shown in Figure~\ref{fixation_area}, fixation spatial coverage was largest for fully sighted participants, narrowest for fully blind participants, with blind and low vision groups falling in between.

Games--Howell tests revealed significant differences between fully blind and fully sighted ($\Delta M = -28.21\%$, $p < 0.001$, Hedges' $g = -3.27$), fully blind and low vision ($\Delta M = -19.50\%$, $p < 0.001$, $g = -1.94$), and blind and fully sighted ($\Delta M = -17.32\%$, $p = 0.04$, $g = -1.62$). A smaller but still significant difference was also observed between low vision and fully sighted ($\Delta M = -8.71\%$, $p = 0.04$, $g = -0.85$). In contrast, no reliable differences were found between blind and fully blind ($p = 0.22$) or between blind and low vision ($p = 0.43$).

Brunner--Munzel tests supported these findings, confirming significant differences between fully blind and low vision ($BM = 7.76$, $p < 0.001$), fully blind and fully sighted ($BM = 75.66$, $p < 0.001$), blind and fully sighted ($BM = 4.35$, $p < 0.001$), and low vision and fully sighted ($BM = 2.81$, $p = 0.01$). No significant differences were found between fully blind and blind ($p = 0.11$) or between blind and low vision ($p = 0.21$).

Together, these results indicate a clear gradation in fixation spatial coverage, with the extent of coverage decreasing systematically as visual function worsens, and the most restricted coverage observed in the fully blind group.

\subsubsection{Individual Fixation Area Patterns}

\begin{figure}[h]
    \centering
    \includegraphics[width=14cm]{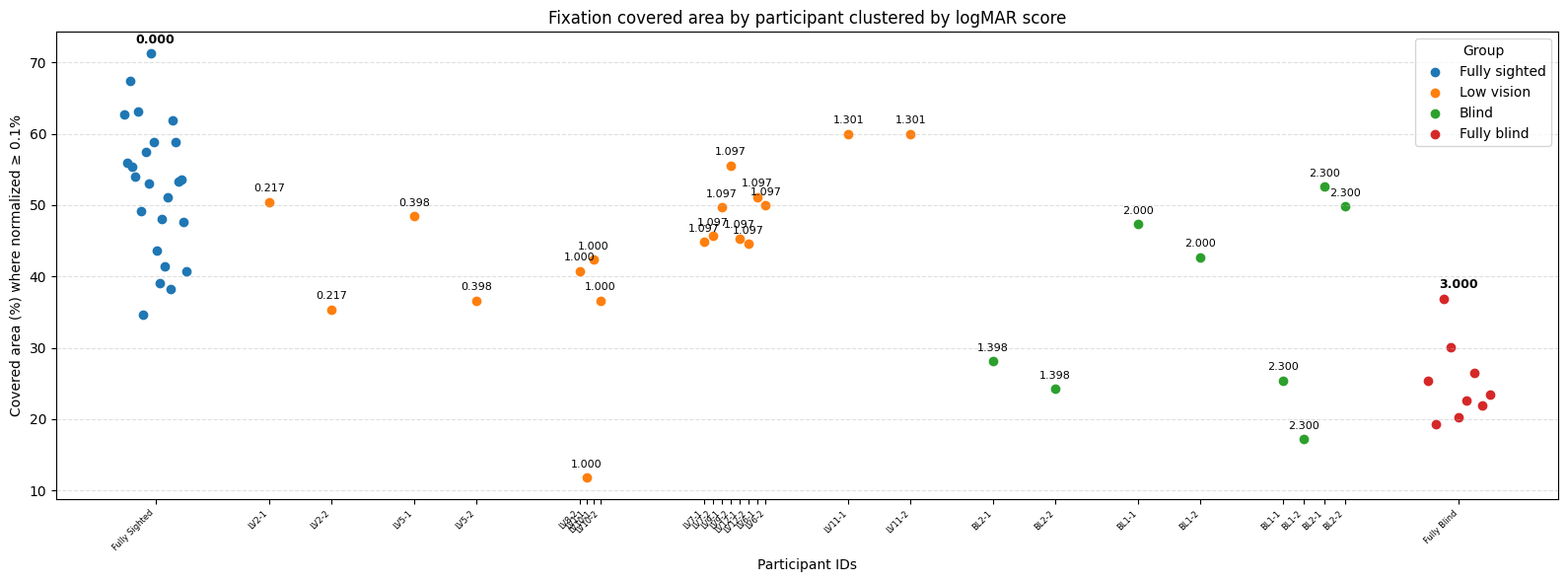}
    \caption{Participant-level fixation area coverage. Each dot represents one participant and trial, grouped by vision category.}
    \label{fixation_count_area}
\end{figure}

Inspection of individual distributions (Figure~\ref{fixation_count_area}) reinforces these group-level trends. Fully sighted participants generally displayed broader and more consistent coverage, suggesting a uniform exploration of the visual scene. In contrast, fully blind participants consistently showed narrow coverage, reflecting limited distribution of fixations across the environment.  

Participants with low vision exhibited greater variability in coverage. Most values overlapped with the fully sighted group, though their upper limits were lower. The blind group showed a mixed pattern: some participants resembled the fully blind group, while others achieved wider coverage.  

Coverage was lowest in the fully blind group, with no participant exceeding 40\%. In contrast, about half of the blind group exceeded this threshold. Most fully sighted and low vision participants exceeded 40\%.

This gradient highlights how residual visual function influences the extent of spatial exploration during navigation.

\subsection{Fixation Direction}

Fixation direction was analyzed separately in terms of azimuth and elevation. Group-level averages and variances are reported in Table~\ref{tab:fixdir}.

\begin{table}[h]
\centering

\label{tab:fixdir}
\begin{tabular}{lcc}
\toprule
Group & Azimuth ($^\circ$) & Elevation ($^\circ$) \\
\midrule
Fully sighted & $-3.33$ (7.17)    & $1.62$ (29.63) \\
Low vision    & $-0.75$ (30.30)   & $0.54$ (61.63) \\
Blind         & $-4.52$ (78.08)   & $10.54$ (72.01) \\
Fully blind   & $-5.47$ (182.04)  & $-4.42$ (22.96) \\
\bottomrule
\end{tabular}
\caption{Group-level means and variances of fixation direction (in degrees). For each participant trial, we computed the average fixation azimuth and elevation. These values were then averaged across all trials within a group to obtain the reported group means. The values in parentheses indicate the variance across trials, reflecting how much fixation direction varied within each group.}

\end{table}

For azimuth, Games--Howell pairwise comparisons revealed no significant differences (all $p > 0.28$), and Brunner--Munzel tests likewise confirmed no significant effects (all $p > 0.13$). Nonetheless, variance increased substantially from the fully sighted to the low vision group, and was largest in the blind and fully blind groups, indicating more dispersed horizontal fixation patterns even though average directions remained near the scene center.

For elevation, Games--Howell tests indicated that the fully blind group differed significantly from both the fully sighted ($\Delta M = -6.04^{\circ}$, $p = 0.03$, $g = -1.12$) and blind groups ($\Delta M = -14.96^{\circ}$, $p = 0.01$, $g = -2.10$). Brunner--Munzel tests confirmed these findings, showing significant differences for fully blind vs.\ fully sighted ($p = 0.004$), fully blind vs.\ blind ($p < 0.001$), fully sighted vs.\ blind ($p = 0.03$), and low vision vs.\ blind ($p = 0.01$). Other contrasts did not reach significance. Importantly, variance was greatest in the blind group, suggesting heterogeneous vertical scanning strategies, whereas the fully blind group showed a more consistent but downward-shifted fixation pattern.

Taken together, these results indicate that although the mean fixation azimuth did not differ significantly across groups, the variability of horizontal scanning increased with vision loss. For elevation, both blind and fully blind groups showed a directional bias. These complementary shifts highlight group-specific strategies in allocating gaze within the scene.

\subsection{Peak Fixation Location}

\begin{figure}[h]
    \centering
    \includegraphics[width=14cm]{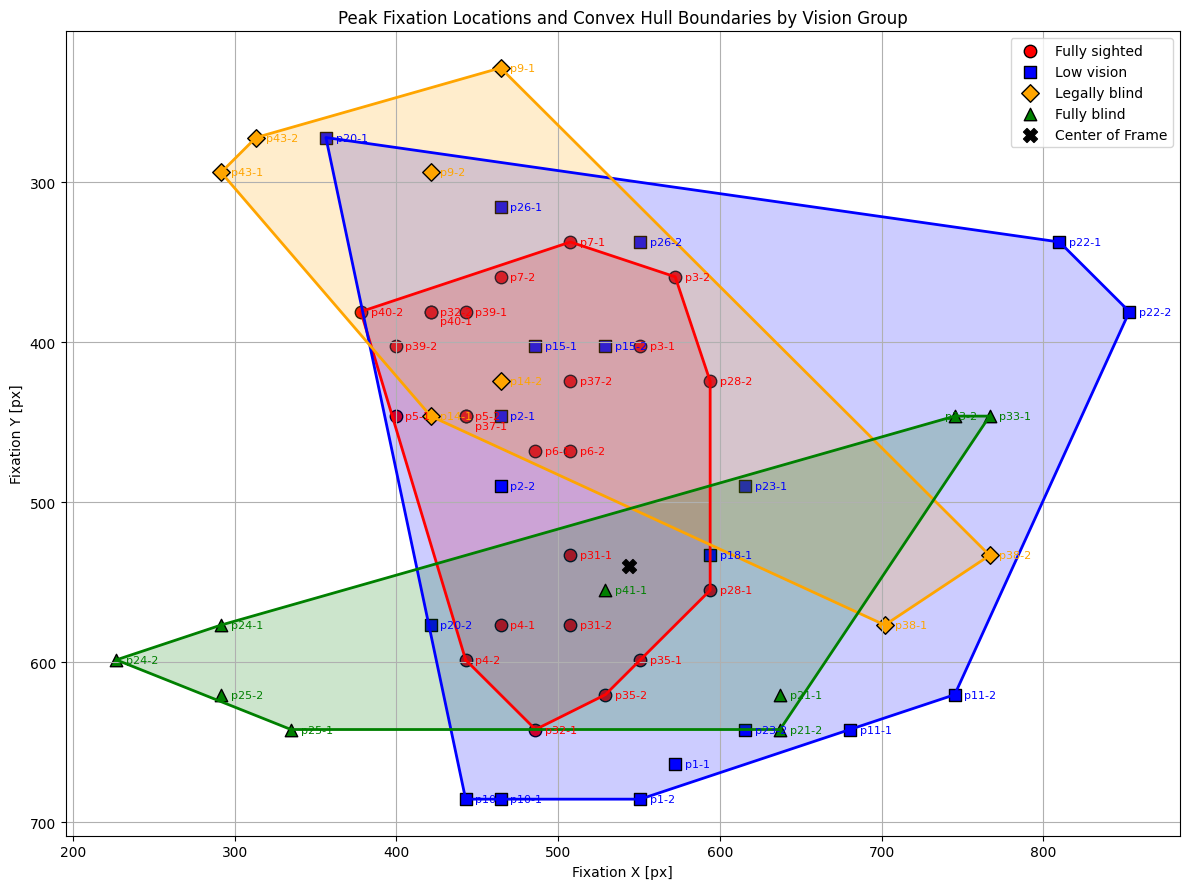}
    \caption{Heatmaps showing peak fixation locations per group. Fully sighted participants concentrated peaks near the center, whereas low vision and blind participants exhibited more dispersed peaks away from center.}
    \label{fig: peak_score}
\end{figure}

As shown in Fig.~\ref{fig: peak_score}, fully sighted participants concentrated peak fixations near the scene center, whereas low vision, blind, and fully blind participants exhibited more dispersed peak locations. 

The mean distance of peak fixation from the scene center was lowest in fully sighted participants ($M = 134.02$ px, variance $= 3365.80$), followed by low vision ($M = 176.22$ px, variance $= 7072.52$), blind ($M = 247.89$ px, variance $= 8078.29$), and fully blind participants ($M = 202.33$ px, variance $= 8414.35$). These larger variances underscore more heterogeneous peak-fixation strategies among vision-impaired groups compared with the consistently center-focused pattern in fully sighted participants.

A Kruskal--Wallis test confirmed significant group differences, $H = 12.38$, $p = 0.01$, $\epsilon^2 = 0.16$. Pairwise Mann--Whitney comparisons with Holm correction showed that fully sighted participants differed significantly from blind participants ($p = 0.02$, $r = 0.71$). The contrast between fully sighted and fully blind participants did not survive correction ($p = 0.07$, $r = 0.57$), nor did comparisons between low vision and blind ($p = 0.21$) or between low vision and fully blind ($p = 0.64$). No significant difference was observed between blind and fully blind participants ($p = 0.64$).

Together, these results indicate that peak fixations in fully sighted participants were concentrated near the center, whereas vision-impaired groups showed more variable and displaced fixation peaks, with blind participants particularly distinguished by greater distances from the scene center.

\subsection{Fixation Area Similarity}

\begin{figure}[h]
    \centering
    \includegraphics[width=14cm]{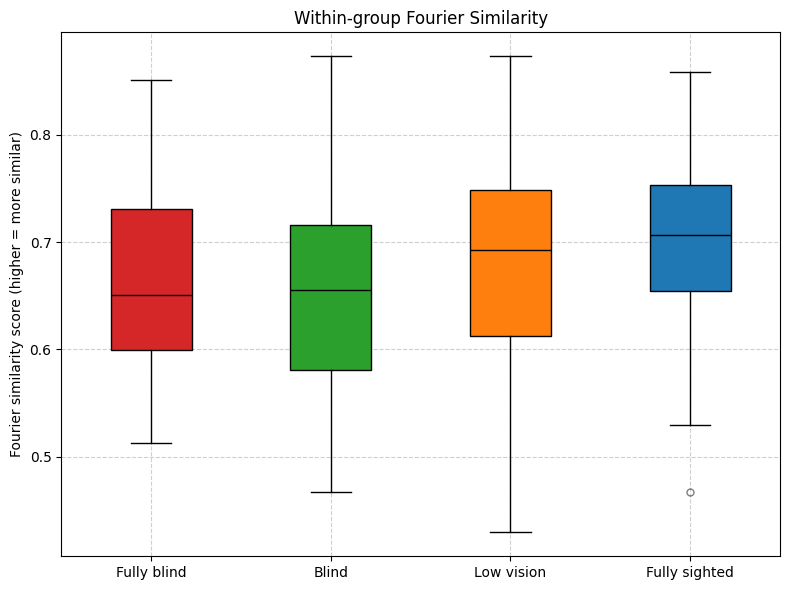}
    \caption{Distribution of Fourier similarity scores across groups. }
    \label{fourier_score}
\end{figure}

Mean Fourier similarity scores were lowest in fully blind participants ($M=0.6600$), 
slightly higher in blind participants ($M=0.6686$), higher in low vision participants ($M=0.6815$), 
and highest in fully sighted participants ($M=0.7046$). 
As illustrated in Fig.~\ref{fourier_score}, higher scores in the fully sighted group indicate more consistent fixation patterns across participants, 
whereas the lower scores in the visually impaired groups reflect greater variability and more individualized fixation strategies.

A Kruskal--Wallis test confirmed an overall effect of group, $H = 13.48$, $p = 0.004$, $\epsilon^2 = 0.02$. Post hoc Dunn tests with Bonferroni correction showed that fully sighted participants had significantly higher similarity than fully blind participants 
(adjusted $p = 0.03$). All other pairwise comparisons were not significant after correction 
(fully sighted vs.\ blind: $p = 0.25$; fully sighted vs.\ low vision: $p = 0.12$; blind vs.\ low vision: $p = 1.00$; 
fully blind vs.\ blind: $p = 1.00$; fully blind vs.\ low vision: $p = 0.83$).

These results indicate that fixation patterns are most uniform in the fully sighted group, while both fully blind and blind participants exhibit more heterogeneous patterns, with low vision intermediate between sighted and blind participants.

\section{Discussion}

This study examined fixation behavior during real-world navigation across four groups: fully sighted, low vision, blind, and fully blind. The results demonstrate that outdoor navigation behavior scales systematically with visual status.  

Overall, walking speed declined steadily with impairment, while fixation counts increased as residual vision decreased. Spatial coverage narrowed progressively from fully sighted to fully blind, and fixation directions became more variable in impaired groups. Peak fixation locations shifted farther from the scene center with increasing impairment, and similarity analyses confirmed less convergence of gaze patterns among vision-impaired participants. Together, these findings indicate a consistent gradient: as visual function worsens, navigation becomes slower, gaze becomes denser but less spatially distributed, and fixation strategies diverge across individuals.

\subsection{Navigation Metrics}

Walking speed declined systematically with low vision: fully sighted participants walked the fastest, followed by those with low vision, blind and then fully blind. This pattern reflects a fundamental tradeoff, as impaired participants slow down to allow more time for environmental sampling. Reduced pace increases the likelihood of detecting near-field hazards and maintaining safety, even at the expense of efficiency. Interestingly, fixation rate is comparable between fully sighted and low vision groups, indicating that although participants with low vision walked more slowly, their visual systems remained just as actively engaged as those of fully sighted individuals. Thus, slower walking speed should not be viewed solely as a performance deficit but rather as a deliberate adaptation to the uncertainty imposed by vision loss.

Participant-level analyses reinforced this interpretation, revealing a clear gradient of decreasing walking speed with more profound vision loss. These findings align with prior work showing that low vision negatively impacts walking speed \cite{hallemans2010low}, further supporting the view that mobility strategies scale progressively with functional vision.

\subsection{Fixation Rate and Fixation Count}

Fixation rate, measured as fixations per minute, showed partial overlap across groups but revealed important differences. Blind participants exhibited significantly higher rates than those with low vision, while fully blind participants also trended higher. This pattern suggests that individuals with more severe vision loss may work harder to extract limited information, engaging in rapid sampling to compensate for uncertainty. In contrast, the low vision group showed the lowest average fixation rate but high variability, possibly reflecting heterogeneous strategies: some participants may have adopted effortful scanning similar to blind participants, whereas others may have restricted fixations to a narrower region immediately in front of them. Fully sighted participants’ fixation rates were not statistically different from the low vision group but showed lower variability. Further studies are needed, as fully sighted participants may scan more in unfamiliar environments to explore, but once familiar, may require fewer fixations. More research is needed to better understand how fixation rate changes with environmental familiarity.

In terms of fixation count, both blind and fully blind participants produced markedly higher totals than fully sighted and low vision groups. This difference arose from their slower walking speeds, which lengthened trial duration, as well as their tendency to interleave repeated checks with exploratory fixations. These elevated counts indicate greater cumulative oculomotor effort for participants with more severe vision loss. Over extended navigation, such increased demand may contribute to fatigue, attentional costs, and reduced responsiveness to hazards. While fixation count captures part of this burden, future work incorporating measures such as fixation duration variability, pupil dilation, or subjective workload could provide a more comprehensive account of the cognitive and physiological demands of navigation under impaired vision.

\subsection{Fixation Rate Fluctuations}

Examining fluctuations in fixation rate across a trip provides insight into participants’ ability to dynamically regulate gaze. The variance pattern observed—highest in fully sighted participants, intermediate in low vision, and lowest in blind participants—suggests that blindness constrains the capacity to modulate fixation rate. Blind participants maintained consistently elevated rates across time, with little opportunity to reduce sampling even during less demanding segments of the route. This “locked-in” profile is consistent with the idea that sustaining a high fixation rate is necessary to compensate for uncertainty in the absence of reliable visual input. In contrast, fully sighted participants exhibited the greatest variability, reflecting the ability to flexibly scale fixation behavior. For example, they may suppress fixations when the environment appears simple or predictable, and increase them when monitoring hazards or complex features becomes necessary. Low vision participants demonstrated intermediate variance, consistent with heterogeneous adaptations shaped by residual vision. Some individuals may have adopted strategies resembling blind participants, sustaining high rates throughout, while others may have intermittently reduced their sampling, more closely resembling sighted behavior.

These results underscore that not only the overall level of fixation activity but also its adaptability is strongly shaped by visual status. Blindness enforces a high and stable fixation regime, sighted individuals retain flexibility to economize effort when possible, and low vision produces diverse but less consistent patterns of modulation. These findings highlight the importance of considering temporal dynamics of fixation behavior, beyond static averages, to better capture the adaptive demands of navigation under varying degrees of vision loss.

\subsection{Spatial Coverage and Fixation Density}
Fixation coverage differed strongly across groups. Fully sighted participants distributed fixations broadly across the visual scene, consistent with a strategy of wide situational monitoring. By contrast, blind and low vision participants concentrated fixations into narrower regions. Importantly, in the low vision group this narrower coverage combined with higher fixation counts, producing denser fixation maps. This pattern suggests that participants with low vision repeatedly sampled specific portions of the scene, strategically directing their residual vision toward areas most relevant for navigation.
Such targeting reflects a well-documented compensatory strategy in individuals with central vision loss, who often adopt a preferred retinal locus (PRL) to replace the damaged fovea \cite{crossland2011preferred}. The PRL acts as a functional substitute, allowing people to place critical features onto a more reliable eccentric retinal area. By anchoring fixations in this way, even degraded visual fields can support effective information extraction, particularly when directed at navigation-relevant landmarks like curbs, doorways, or obstacles. The denser coverage observed here is consistent with this mechanism: repeated sampling of the same region increases the likelihood of detecting salient cues, reduces uncertainty, and extends the utility of limited visual input. More broadly, these findings underscore how residual vision is not merely a diminished resource but one that can be optimized through strategic allocation, effectively increasing the impact of each fixation in real-world mobility.

\subsection{Fixation Patterns and Peak Locations}

Fixation behavior revealed a fundamental asymmetry between groups. Fully sighted participants converged on a shared, low-variance, center-focused template that economizes eye movements and supports stable horizon and path monitoring. This template often manifested as a T-shaped distribution, characterized by horizontal sampling of the horizon and vertical checks of the walking path, with occasional fixations near the right front of the feet. Notably, similar T-shaped arrangements have been reported in other contexts, such as face perception \cite{Greene2022EmotionRecognition} and fixation-based tasks \cite{corn2010foundations}, suggesting that this organization may represent a general principle of human oculomotor control rather than a navigation-specific artifact.

In contrast, participants with visual impairments showed pronounced heterogeneity, and this variability is not merely superficial. Blind participants exhibited narrowly restricted circular fixation clusters, reflecting minimal responsiveness to environmental features. Low vision participants displayed highly diverse patterns shaped by the location and quality of their residual vision. Crucially, this heterogeneity should be interpreted along a spectrum from adaptive to maladaptive responses. On the adaptive end, some participants reorganized their sampling toward more informative regions, such as greater peripheral scanning when central vision was compromised. On the maladaptive end, others engaged in inefficient or fragmented scanning: elevated saccade rates with shortened fixations, frequent corrective or back-tracking saccades, oversampling of scotoma-adjacent regions, or narrow looping searches that left large areas of the scene unsampled. Such behaviors risk “maxing out” the oculomotor budget within a trial, inflating cognitive and oculomotor load and ultimately contributing to fatigue and reduced navigation efficiency.

Fourier-based shape similarity analyses corroborated these observations. Fully sighted participants exhibited the highest within-group similarity, confirming their convergence on a shared fixation template. By contrast, both low vision and blind groups had markedly lower similarity scores compared with the fully sighted group, consistent with the absence of a unifying strategy. Importantly, lower similarity here reflects not just diversity but unequal adaptation: some individuals developed efficient compensations, while others relied on maladaptive patterns that increased effort without yielding commensurate perceptual benefit.

Taken together, these findings highlight that heterogeneity in the visual disability groups has functional consequences. It represents a mix of adaptive and maladaptive scanning strategies, with the latter imposing excessive eye-movement demands that likely accelerate fatigue, compromise hazard detection, and hinder navigation. Recognizing this distinction underscores the need for targeted interventions—such as gaze-strategy training, cueing to promote forward preview, or assistive feedback to reduce redundant revisits—that can shift scanning from maladaptive toward more efficient patterns.

\subsection{High Variability in the Low Vision Group}
Across nearly all metrics—including walking speed, fixation counts, fixation rate, fixation area, peak fixation locations, and distance to the scene center—the low vision group exhibited the greatest variability. While many of these measures showed a general positive correlation with visual acuity, the low vision participants produced numerous outliers, indicating that their strategies were less homogeneous than those of the fully sighted or blind groups. Some participants with relatively poor acuity exhibited patterns resembling the fully sighted group, potentially reflecting benefits from rehabilitation training or accumulated navigation experience. These findings highlight that residual vision is used in highly individualized ways, shaped by factors beyond acuity alone.

This variability likely reflects the diverse circumstances and adaptations within the low vision population. Differences in age, level of education, and capacity to learn and apply new rehabilitation techniques all contribute to how effectively individuals can make use of residual vision. Even when participants receive similar rehabilitation training, their ability to translate these techniques into consistent fixation behavior varies considerably. Some individuals may adopt highly effective strategies, resulting in performance that approaches the fully sighted group, while others may struggle to implement the same methods and show less stable outcomes.

Importantly, this heterogeneity also arises from differences in the underlying pathology and severity of impairment. Conditions such as macular degeneration, retinitis pigmentosa, and diabetic retinopathy each alter the visual field in distinct ways, shaping how residual vision can be deployed during navigation. These distinctions suggest that a “one-size-fits-all” approach may be insufficient: optimizing mobility outcomes may require tailoring fixation and scanning strategies to the specific characteristics of a given condition, its severity, and related mobility metrics. In this sense, the variability observed in the low vision group highlights both the individualized nature of rehabilitation and the need for condition-specific strategies that adapt gaze behavior to maximize the functional use of residual vision.

\subsection{Implications for Rehabilitation and Assistive Technology}

Fixation rate appeared to increase as vision worsened, with blind participants already operating at very high rates that remained elevated across contexts. This pattern suggests that fixation rate in visual disability may be pinned close to the physiological ceiling, a state that likely imposes cognitive and oculomotor costs and contributes to fatigue. Rather than targeting fixation rate directly, rehabilitation should focus on improving the allocation of these already frequent fixations.

A promising approach is to guide participants toward strategies that broaden scene coverage and reduce inefficient re-checking, thereby improving efficiency without increasing total eye movements. Pairing rehabilitation with real-time eye tracking could provide immediate feedback, allowing therapists to highlight adequate scanning behaviors, identify maladaptive patterns such as excessive revisits or missed forward preview, and coach more effective allocation. Structured strategies—such as forward-oriented scanning to anticipate hazards—may help participants harness their existing fixation capacity more productively, mitigating fatigue while enhancing safety.

The high variability observed in the low vision group highlights the need for individualized approaches. Unlike the fully sighted group, which exhibited a consistent center-focused, T-shaped fixation strategy, low vision participants relied on heterogeneous behaviors shaped by the location of residual vision, duration of impairment, prior training, and personal adaptation. This variability underscores that standardized solutions may be ineffective. Rehabilitation and assistive technologies must be tailored to the individual’s specific visual profile, emphasizing the unique strengths and limitations of their residual vision.

Slower walking speed, consistently observed with greater vision loss, should not be regarded solely as a deficit. By moving more slowly, participants accumulated more fixations across the same route, thereby increasing the total information gathered from the environment. This deliberate tradeoff—reduced efficiency in exchange for enhanced safety—illustrates how mobility and visual exploration strategies are tightly coupled. Rehabilitation can build on this adaptive behavior by helping participants allocate their additional fixations more strategically—for example, emphasizing forward-looking areas that support hazard anticipation. In some contexts, it may even be beneficial to encourage intermittent pauses for deliberate visual sampling, allowing individuals to gather information more stably before resuming movement, rather than relying solely on rapid fixations while in motion.

For assistive technology, the goal should be to complement, rather than duplicate, the user’s existing perceptual abilities. Devices that highlight peripheral cues may benefit individuals with central vision loss, while those with peripheral field loss may require enhanced central detail. Real-time gaze monitoring offers particular promise: by detecting maladaptive behaviors such as excessive re-checking or upward fixation, adaptive systems could provide gentle corrective prompts, reducing both visual and cognitive load during navigation.

Finally, the structured T-shaped fixation patterns consistently observed in fully sighted participants point to potential training targets. Similar distributions have been reported in controlled studies of face viewing and other visual tasks, suggesting this may be a stable and efficient visual exploration template. Teaching visually impaired individuals to adopt systematic forward-oriented scanning patterns may therefore enhance spatial awareness and obstacle anticipation. Coupled with periodic gaze monitoring to track progress, such approaches could provide therapists with objective tools to refine training protocols and improve outcomes. Taken together, these insights emphasize that effective rehabilitation and assistive technologies must leverage both biological constraints and adaptive strategies, ultimately supporting safer and more efficient real-world navigation.

\subsection{Future Directions}

Future research should build on these findings by examining how fixation strategies evolve over time with rehabilitation training or increased experience with assistive devices. Longitudinal studies would clarify the adaptability of gaze behavior and the long-term effectiveness of interventions designed to improve navigation skills. Identifying which strategies are most effective for specific impairments could guide rehabilitation programs toward promoting the most beneficial approaches and maximizing the use of residual vision.

The diversity of environments in which navigation occurs also warrants closer study. Beyond controlled outdoor routes, testing should extend across a range of real-world settings, including indoor hallways, transit hubs, intersections, construction zones, and crowded public spaces. Environmental factors such as low light, high crowd density, obstacle frequency, and poor weather may strongly influence fixation strategies in both fully sighted and visually impaired individuals. Understanding these adjustments would provide a more complete picture of navigation behavior and could inform not only the design of safer, more accessible environments but also the refinement of rehabilitation training. Assistive technologies could likewise adapt to contextual demands by prompting users to modify their gaze strategies: for example, encouraging longer fixations on high-contrast landmarks in low light, quick forward-oriented checks in crowded areas, or downward-directed sampling in cluttered obstacle fields. Such context-sensitive guidance would maximize perception and safety across diverse real-world settings.

Future studies should also expand the set of dependent variables beyond gaze alone. Navigation is inherently multimodal, and spatiotemporal gait parameters such as walking speed, stride length, step variability, pause frequency, turning angles, and obstacle clearance offer critical insight into how visual exploration translates into mobility outcomes. Coupling eye-tracking with gait analysis would allow researchers to identify how maladaptive fixation patterns manifest in inefficient movement, and conversely, how compensatory strategies support safer and more efficient locomotion.

Some participants with blindness or low vision already adopt strategies that enhance their functional performance, such as slowing walking speed to allow more fixations and gather additional environmental information. Future studies should aim to isolate which compensatory strategies are most effective and explore how they can be systematically taught. This would ensure that remaining vision is used as efficiently as possible, ultimately improving independence and safety.

Finally, integrating gaze monitoring into rehabilitation could provide an objective means of tracking progress and tailoring interventions. Periodic assessment of fixation patterns would allow clinicians to identify maladaptive behaviors—such as overly narrow scanning or upward-directed fixations—and introduce targeted exercises to promote more effective strategies. By focusing on individualized needs and tracking change over time, future work can support the development of interventions that are both more effective and more sustainable.
.

\subsection{Limitations}

This study has several limitations that should be acknowledged. First, the sample size was relatively small, which limits the statistical power and generalizability of the findings. Larger and more diverse cohorts will be needed to validate these results across broader populations of individuals with low vision and blindness.

Second, the accuracy and robustness of the eye-tracking hardware were limited. The Pupil Invisible device achieved an average gaze accuracy of only $4.6^{\circ}$ in uncalibrated mode, which is lower than the precision of laboratory-based systems. In addition, pupillometry data (e.g., pupil diameter, eye state) were not available, the scene video resolution was relatively low ($1088 \times 1080$ px at 30 Hz), the field of view was restricted (H: $82^{\circ}$, V: $82^{\circ}$), and the tethered connection to a smartphone frequently disconnected during motion. Together, these constraints may have reduced measurement accuracy, ecological validity, and robustness in real-world outdoor conditions.

Third, this study did not analyze how fixations corresponded to specific environmental stimuli. For example, whether participants with different vision statuses fixated differently when encountering pedestrians, scanning for traffic lights, or avoiding static obstacles on the street remains unknown. These distinctions are critical for understanding the functional role of gaze in navigation and should be addressed in future research.

\section{Conclusion}

This study examined gaze during outdoor navigation across fully sighted, low vision, and blind participants. Fully sighted individuals converged on a consistent, center-focused T-shaped fixation template, while fully blind participants showed highly restricted fixation areas and produced the highest overall fixation counts. Low vision participants occupied an intermediate position but with the greatest within-group variability, reflecting heterogeneous use of residual vision.  Blind group also occupied an intermediate position but behaved more similar to fully blind group.

Fixation rate was not statistically different between fully sighted and low vision groups, but the combination of slower walking speeds and elevated fixation counts in the vision-impaired groups points to greater oculomotor demand. Blind and fully blind groups have higher fixation rate than fully sighted and low vision groups, indicating blind participants might work harder in using their residual vision. These findings emphasize that gaze strategies are vision-dependent but also reflect unequal adaptation: some individuals develop efficient compensations, while others rely on maladaptive patterns that increase fatigue. Effective rehabilitation and assistive technology must therefore move beyond acuity alone, tailoring interventions to functional gaze use in real-world contexts and promoting more efficient, sustainable scanning strategies.

\section{Acknowledgment}
This research was supported by the National Science Foundation under Grant Nos. ITE-2345139, CNS-1952180,
and ITE-2236097 by the National Eye Institute and Fogarty International Center under Grant No. R21EY033689, as
well as by the U.S. Department of Defense under Grant No. VR200130. The content is solely the responsibility of the authors and does not necessarily represent the official views of the National Institutes of Health, National Science Foundation,
and Department of Defense.

This manuscript employed ChatGPT-5 (OpenAI) for grammar checking and language refinement, while ChatGPT-5 was utilized for generating codes for data analysis and visualization. The authors carefully reviewed and validated all outputs to ensure accuracy and maintain the research’s intellectual rigor.

\section{Disclosure statement}
No potential conflict of interest was reported by the author(s).

\FloatBarrier
\clearpage

\bibliographystyle{ieeetr}   
\bibliography{references}    

\FloatBarrier
\end{document}